\documentclass[10pt,a4paper,twosides]{amsart}

\usepackage{stmaryrd}
\usepackage{amsmath}
\usepackage{amssymb}
\usepackage{tabularx}
\input psfig.sty
\usepackage[top=2.54cm,bottom=3.54cm,left=3.15cm,right=3.15cm ]{geometry}

\newcommand{\Vext}{V_{\rm ext}}
\newcommand{\Vp}{V_P}
\newcommand{\Cp}{C_P}

\newcommand{\p}{\partial}

\newcommand{\fl}[2]{\frac{#1}{#2}}

\newcommand{\btd}{\nabla}

\newcommand{\be}{\begin{equation}}
\newcommand{\ee}{\end{equation}}
\newcommand{\ba}{\begin{array}}
\newcommand{\ea}{\end{array}}
\newcommand{\bea}{\begin{eqnarray}}
\newcommand{\eea}{\end{eqnarray}}
\newcommand{\beas}{\begin{eqnarray*}}
\newcommand{\eeas}{\end{eqnarray*}}

\newtheorem{lemma}{Lemma}[section]

\newcommand{\bx}{{\bf x} }

\begin{document}

\title[Simplified pseudospectral method for spherically symmetric SPS]{A short note on simplified pseudospectral methods for computing ground state
and dynamics of spherically symmetric Schr\"{o}dinger--Poisson--Slater system}

\date{\today}
\author[X. Dong]{Xuanchun Dong}

 \address{{\bf Xuanchun Dong}\newline
 Center for Computational Science and Engineering\newline
          Department of Mathematics\\
          National  University of Singapore\newline
          Block S17, 10, Lower Kent Ridge Road, 119076, Singapore}
 \email{dong.xuanchun@nus.edu.sg\\
        dong.xuanchun@gmail.com}


\begin{abstract}
In a recent paper we proposed and compared various approaches to compute the ground state and dynamics of the Schr\"{o}dinger--Poisson--Slater (SPS) system for general external potential and initial condition, concluding that the methods based on sine pseudospectral discretization in space are the best candidates.  This note is concerned with the case that the external potential and initial condition are spherically symmetric.  For the SPS system with spherical symmetry, via applying a proper change of variables into the reduced quasi-1D model we simplify the methods proposed for the general 3D case such that both the memory and computational load are significantly reduced.

\end{abstract}



\maketitle

\section{Introduction}

The Schr\"{o}dinger--Poisson--Slater (SPS) system, or the
Schr\"{o}dinger--Poisson--X${\alpha}$ system, serves as a local single
particle approximation of the time-dependent Hartree--Fock equations.
For its formal derivation we refer the readers to \cite{Mauser,ZhangDong} and references therein.
The SPS system reads, in scaled form, \begin{align} \label{spsorig} & i \p_t
 \psi(\bx,t)= -\fl{1}{2} \btd^2  \psi +\Vext(\bx) \psi +
 \Cp  \Vp(\bx,t) \psi  -\alpha  |\psi|^{\fl{2}{3}} \psi,\quad \bx \in{\Bbb R}^3,\quad t>0,\\
 & \label{spsorig2} -\btd^2 \Vp(\bx,t)  = |\psi |^2,\quad \bx \in{\Bbb R}^3,\quad t\geq 0\\
 & \label{spsorig3}\psi(\bx,t=0)=\psi_0(\bx),\quad \bx \in{\Bbb R}^3.
\end{align}
Here, the complex--valued $\psi(\bx, t)$ is the
single particle wave function with $\lim_{|\bx|\to
\infty}|\psi(\bx,t)|= 0$ exponentially fast, $\Vext(\bx)$ is a given external
potential, for example a confining potential, $\Vp(\bx, t)$ stands for
the Hartree potential with decay condition $\lim_{|\bx|\to
\infty} \Vp(\bx,t)=0$, and $\Cp$ ($\Cp>0$ for
repulsive interaction and $\Cp<0$ for attractive interaction) and
$\alpha$ ($\alpha>0$ for electrons due to the physical nature) are interaction constants.

 There exit at least two important invariants of
the SPS system (\ref{spsorig})-(\ref{spsorig3}): the normalized {\sl mass}
 \be\label{norm}
 N\left(\psi(\cdot,t)\right):=\left\|\psi(\cdot,t)\right\|^2= \int_{{\Bbb R}^3}
\left|\psi(\bx,t)\right|^2\,d \bx=1,\quad t\geq 0, \ee
and the {\sl energy} \be \label{energy}
E(\psi(\cdot,t)):=\int_{{\Bbb R}^3}\left[ \frac{1}{2}\left|\nabla\psi
\right|^2 +\left(\Vext(\bx)
  +\fl{\Cp}{2}\Vp(\bx,t)\right)|\psi |^2-
 \fl{3 \alpha}{4}|\psi |^{\fl{8}{3}} \right]d \bx,\quad t\geq 0. \ee
And, the ground state is defined as the minimizer of the following constraint minimization problem:

\noindent Find $\phi_g\in S=\left\{\phi \,|\, E(\phi)<\infty,\,N\left(\phi\right)=1\right\}$ such that
\begin{equation}\label{min_prob}
E_g:=E(\phi_g)=\min_{\phi\in S} E(\phi).
\end{equation}

There is a series of analytical results on the SPS system in literatures; see, e.g., \cite{BGM,Masaki,Stimming} for its well-posedness and \cite{BLS,Choquard1,Lieb,Soler} for the existence and uniqueness of its ground states.  Also, a detailed review was given in \cite[Section 1]{ZhangDong}.
On the other hand, the numerics of the SPS system was considered in, e.g., \cite{BMS,ADM,CLLM,EZ,TodNum2,ZhangDong}.  In particular, in \cite{ZhangDong} we proposed and compared different methods to compute the ground state and dynamics of the SPS system for general external potential and initial condition, ending with a conclusion that a backward Euler sine pseudospectral (BESP) method and a time-splitting sine pseudospectral (TSSP) method are the best choices to approximate the ground state and dynamics respectively.  However, we have pointed out that when the external potential and initial conditions are with spherical symmetry, the original 3D problem reduces to a quasi-1D problem, for which the spectral-type methods BESP and TSSP cannot be directly extended and we suggested to apply the standard finite-difference to space derivatives \cite[Remarks 3.2 and 4.1]{ZhangDong}.  The objective of this note is to propose spectral-type methods for the spherically symmetric case, which simplify the BESP and TSSP methods proposed in \cite{ZhangDong} for general 3D case, with the help of a proper change of variables for the reduced quasi-1D model.  The simplified methods are still spectrally accurate in space, but reduce the memory cost from $O(J^3)$ to $O(J)$ and the computational cost per time step from $O(J^3\ln(J^3))$ to $O(J\ln(J))$, where $J$ is the number of mesh nodes.

The rest is organized as follows.  In Section 2 we give a reduced quasi-1D model for the spherically symmetric case.  In Section 3 simplified BESP and TSSP methods are proposed and in Section 4 numerical results are reported.  Finally, some concluding remarks are drawn in Section 5.

\section{A quasi-1D model reduced from spherically symmetric system}
 \setcounter{equation}{0}

Throughout this note, we assume that both the external potential $\Vext$ and initial condition $\psi_0$ are spherically symmetric, i.e., $\Vext(\bx)=\Vext(r)$ and $\psi_0(\bx)=\psi_0(r)$ with $r=|\bx|$.  In this case, the solution $\psi$ of (\ref{spsorig})-(\ref{spsorig3}) and the ground state $\phi_{g}$ are also spherically symmetric, i.e.,
\begin{equation*}
\psi(\bx,t) = \psi(r,t),\quad \phi_{g}(\bx)=\phi(r),\quad \bx \in {\Bbb R}^3,\quad t\geq 0.
\end{equation*}
Thus, the SPS system (\ref{spsorig})-(\ref{spsorig3}) collapses the following quasi-1D problem
\begin{align}
\label{1d_model_1}& i \p_t \psi(r,t) = -\frac{1}{2r^2}\frac{\p}{\p r}\left(r^2\frac{\p \psi}{\p r}\right) +\Vext(r)\psi
+ \Cp  \Vp(r,t)\psi - \alpha  \left|\psi\right|^{\frac{2}{3}} \psi,\quad 0< r <\infty, \quad t>0,\\
\label{1d_model_2}& -\frac{1}{r^2}\frac{\p}{\p r}\left(r^2\frac{\p \Vp(r,t)}{\p r}\right) = \left|\psi\right|^2,\quad 0< r <\infty, \quad t\geq 0,\\
\label{1d_model_3}& \psi(r,t=0) = \psi_0(r),\quad 0 \leq  r <\infty,
\displaybreak[1]
\intertext{with boundary conditions}
\label{1d_model_4}& \p_r \psi(0,t)= \p_r \Vp(0,t) = 0,\quad
\lim_{r\to \infty} \psi(r,t)=0,\quad \lim_{r\to \infty} r\Vp(r,t) = \frac{1}{4\pi},\quad t\geq 0,
\end{align}
due to the decay conditions of $\psi$ and $\Vp$, and the Green function of the Laplacian on ${\Bbb R}^3$ \cite{Kellogg}.

Introducing
\begin{equation}
{\mathcal U}(r,t) = 2\sqrt{\pi} r\psi(r,t),\quad {\mathcal V}(r,t) = 4\pi r \Vp(r,t),\quad 0\leq r <\infty, \quad t\geq 0,
\end{equation}
a simple computation shows
\begin{equation}
\frac{1}{r^2}\frac{\p}{\p r}\left(r^2\frac{\p \psi }{\p r}\right) = \frac{1}{2\sqrt{\pi}r}\p_{rr}{\mathcal U},\quad
\frac{1}{r^2}\frac{\p}{\p r}\left(r^2\frac{\p \Vp}{\p r}\right) = \frac{1}{4\pi r}\p_{rr}{\mathcal V}.
\end{equation}
We remark here that the similar technique has been used in \cite{BD2011,EZ}.
Plugging the above into (\ref{1d_model_1})-(\ref{1d_model_4}), we obtain
\begin{align}
\label{1d_model_cv_1}& i\p_t {\mathcal U}(r,t) = -\frac{1}{2}\p_{rr} {\mathcal U} + \Vext(r){\mathcal U}
+ \frac{\Cp}{4\pi r}{\mathcal V}(r,t){\mathcal U}
-{\alpha}{\left(2\sqrt{\pi}r\right)^{-\frac{2}{3}}}\left|{\mathcal U}\right|^{\frac{2}{3}}{\mathcal U},
\quad 0< r <\infty,\quad t>0,\\
\label{1d_model_cv_2}& -\p_{rr}{\mathcal V}(r,t) = \frac{1}{r}\left|{\mathcal U}\right|^2,\quad 0< r <\infty,\quad t\geq 0,\\
\label{1d_model_cv_3}& {\mathcal U}(r,t=0)={\mathcal U}_0(r)=2\sqrt{\pi} r \psi_0(r),\quad 0\leq r < \infty,\\
\label{1d_model_cv_4}& {\mathcal U}(0,t) = {\mathcal V}(0,t)=0,\quad
\lim_{r\to \infty}{\mathcal U}(r,t)=0,\quad \lim_{r\to
\infty}{\mathcal V}(r,t) = 1,\quad t\geq 0.
\end{align}
Also, the above problem conserves the {\sl mass}
\begin{equation*}
{\mathcal N}({\mathcal U}(\cdot,t)):= \left\|{\mathcal U}(\cdot,t)\right\|^2
= \int_0^{\infty} \left|{\mathcal U}(r,t)\right|^2 dr=N(\psi(\cdot,t))=1,\quad t\geq 0,
\end{equation*}
and the {\sl energy}
\begin{equation*}
{\mathcal E}({\mathcal U}(\cdot,t)) : = \int_0^{\infty} \left[ \frac{1}{2}\left|\p_r {\mathcal U}\right|^2
+\left(\Vext(r)+\frac{\Cp}{8\pi r}{\mathcal V}(r,t)\right)\left|{\mathcal U}\right|^2
-\frac{3\alpha}{4} \left(2\sqrt{\pi} r\right)^{-\frac{2}{3}} \left|{\mathcal U}\right|^{\frac{8}{3}} \right] dr
=E(\psi(\cdot,t)),
\quad t\geq 0.
\end{equation*}
In what follows we will take the problem (\ref{1d_model_cv_1})-(\ref{1d_model_cv_4}) as the starting model to propose efficient numerical methods.  After we get the solution ${\mathcal U}$ of (\ref{1d_model_cv_1})-(\ref{1d_model_cv_4}), the solution $\psi$ of (\ref{1d_model_1})-(\ref{1d_model_4}) is obtained as
\begin{equation}
\psi(r,t) = \frac{1}{2\sqrt{\pi}}
\left\{
\begin{array}{ll}
{\mathcal U}(r,t)/r,& r>0,\\
\p_r {\mathcal U}(r,t) = \lim_{s\to 0^{+}}{\mathcal U}(s,t)/s,& r=0,
\end{array}
 \right. \quad t\geq 0.
\end{equation}

Meanwhile, the minimization problem (\ref{min_prob}) to define the ground state collapses to:

\noindent Find $\varphi_g\in {\mathcal S}=\left\{\varphi\, |\, {\mathcal E}(\varphi)<\infty,\,{\mathcal N}(\varphi)=1, \,\varphi(0)=0\right\}$ such that
\begin{equation}\label{gs_cv}
{\mathcal E}_g:={\mathcal E}(\varphi_g) = \min_{\varphi \in {\mathcal S}} {\mathcal E}(\varphi).
\end{equation}
Again, after we get the minimizer of (\ref{gs_cv}), the ground state $\phi_g$ of (\ref{1d_model_1})-(\ref{1d_model_3}) is obtained as
\begin{equation}
\phi_g(r) = \frac{1}{2\sqrt{\pi}}
\left\{
\begin{array}{ll}
\varphi_g(r)/r,& r>0,\\
\p_r \varphi_g(r) = \lim_{s\to 0^{+}}\varphi_g(s)/s,& r=0.
\end{array}
 \right.
\end{equation}

\section{Efficient numerical methods}
\setcounter{equation}{0}


\subsection{Backward Euler sine pseudospectral method for ground state}

Choose a time step $\Delta t>0$ and set $t_n = n\Delta t$ for $n=0,1,\ldots.$ Similar as \cite[Section 2]{ZhangDong}, for the minimization problem (\ref{gs_cv}), we construct the following gradient flow with discrete normalization (GFDN):
\begin{align}
& \p_t \varphi (r,t)  
= \frac{1}{2}\p_{rr}\varphi - \Vext(r)\varphi - \frac{\Cp}{4\pi r} {\mathcal V}(r,t) \varphi
+ {\alpha}{\left(2\sqrt{\pi}r\right)^{-\frac{2}{3}}}\left|\varphi\right|^{\frac{2}{3}}\varphi,
\quad 0<r<\infty,\quad t_n \leq t < t_{n+1},\\
& -\p_{rr}{\mathcal V}(r,t) = \frac{1}{r}\left|\varphi\right|^2,\quad 0< r  < \infty, \quad t\geq 0,\quad
\varphi(r,t_{n+1}):=\varphi(r,t_{n+1}^+)=\frac{\varphi(r,t_{n+1}^-)}{\left\|\varphi(r,t_{n+1}^-)\right\|},\quad n\geq 0,\\
& \varphi(r,t=0) = \varphi_0(r),\quad 0\leq r < \infty,\quad \mbox{with}\quad  {\mathcal N}(\varphi_0)=1,\\
& \varphi(0,t) = {\mathcal V}(0,t)=0,\quad \lim_{r\to
\infty}\varphi(r,t)=0, \quad \lim_{r\to \infty}{\mathcal V}(r,t) =
1,\quad t\geq 0,
\end{align}
where $\varphi(r,t_n^{\pm}):=\lim_{t\to t_n^{\pm}}\varphi(r,t)$ for $0\leq r < \infty$.  In practical computation, we truncate the above problem into an interval $[0,R]$ with $R>0$ sufficiently large, together with Dirichlet boundary conditions
\begin{equation*}
\varphi(0,t)=\varphi(R,t)={\mathcal V}(0,t)=0,\quad {\mathcal V}(R,t) = 1,\quad t\geq 0.
\end{equation*}
Introducing  a linear translation (homogenization) $\overline{\mathcal V}(r,t) = {\mathcal V}(r,t) - r/R$ for $0\leq r \leq R$,
\begin{equation}\label{linear_trans}
-\p_{rr} \overline{\mathcal V }(r,t) =-\p_{rr}{\mathcal V}(r,t) = \frac{1}{r}\left|\varphi\right|^2,\quad 0< r  < R, \quad
 \overline{\mathcal V }(0,t)= \overline{\mathcal V }(R,t) =0,\quad  t\geq 0.
\end{equation}
Then we discretize the problem in space by sine pseudospectral method and in time by a backward Euler integration similar as that used in \cite{ZhangDong}.  Choose a mesh size $h_r=\Delta r = R/J$ with some even integer $J>0$, and denote the grid points as $r_j = j h_r$ for $j=0,1,\ldots,J$.
Let $\varphi_j^n\approx \varphi(r_j, t_n)$ and $\overline{\mathcal V}_j^n \approx \overline{\mathcal V}(r_j,t_n)$, and denote $\rho_j^n = \left|\varphi_j^n\right|^2/r_j$.
Choosing $\varphi_j^0 = \varphi_0(r_j)$, a backward Euler sine pseudospectral discretization (BESP) reads: for $n=0,1,\ldots,$
\begin{align}
\label{besp1}& \frac{\varphi_j^+ - \varphi_j^n}{\Delta t} =
\frac{1}{2}\left.\left(D_{rr}^s\varphi^+\right)\right|_{j}
-\left[\Vext(r_j)+\frac{\Cp}{4\pi r_j}\overline{\mathcal
V}_j^n+\frac{\Cp}{4\pi R}
-{\alpha}{\left(2\sqrt{\pi}r_j\right)^{-\frac{2}{3}}}\left|\varphi_j^n\right|^{\frac{2}{3}}\right]\varphi_j^+,
\quad j=1,2,\ldots,J-1,\\
\label{besp2}& -\left.\left(D_{rr}^s\overline{\mathcal V}^n\right)\right|_{j}  = \rho_j^n,\quad j=1,2,\ldots, J-1,\quad
\varphi_0^+ = \varphi_J^+ = \overline{\mathcal V}_0^n = \overline{\mathcal V}_J^n = 0,\\
\label{besp3}& \varphi_j^{n+1} =
\frac{\varphi_j^+}{\left\|\varphi^+\right\|_h},\quad j=1,2,\ldots,
J-1,\quad \mbox{with}\quad \left\|\varphi^+\right\|_h^2:= h_r
\sum_{j=1}^{J-1}\left|\varphi_j^+ \right|^2,
\end{align}
where $D_{rr}^s$ is the sine pseudospectral approximation of $\p_{rr}$, defined via
\begin{equation}
-\left.\left(D_{rr}^s\varphi^n\right)\right|_{j} = \sum_{k=1}^{J-1}\mu_k^2 \widetilde{\left({\varphi^n}\right)}_k
\sin\left(\frac{jk\pi}{J}\right),\quad j=1,2,\ldots, J-1,
\end{equation}
with $\left(\widetilde{\varphi^n}\right)_k$ the discrete sine transform coefficients
\begin{equation}
\widetilde{\left({\varphi^n}\right)}_k = \frac{2}{J}\sum_{j=1}^{J-1} \varphi_j^n \sin\left(\frac{jk\pi}{J}\right),\quad
\mu_k = \frac{k\pi}{R},\quad k=1,2,\ldots, J-1.
\end{equation}
Similar as \cite{ZhangDong}, the linear system (\ref{besp1})-(\ref{besp3}) can be iteratively solved efficiently in phase space with the help of discrete sine transform.  After we get the stationary solution $(\varphi_g)_j$ of the above problem, the ground state $(\phi_g)_j\approx \phi_g(r_j)$ of
(\ref{1d_model_1})-(\ref{1d_model_3}) is achieved via
\begin{equation}
(\phi_g)_j = \frac{1}{2\sqrt{\pi}}
\left\{
\begin{array}{ll}
(\varphi_g)_j/r_j,& j=1,2,\ldots,J,\\
\sum_{k=1}^{J-1} \mu_k \widetilde{\left(\varphi_g\right)}_k,& j=0.
\end{array}
\right.
\end{equation}
 Note that the above numerical method is spectrally accurate and it works only when $\Vext$ is spherically symmetric.  Compared with the pseudospectral method proposed in \cite{ZhangDong} for general 3D problem, the memory cost is reduced from $O(J^3)$ to $O(J)$ and computational cost per time step is reduced from $O(J^3\ln(J^3))$ to $O(J\ln(J))$.

\subsection{Time-splitting sine pseudospectral method for dynamics}

Again, we truncate the problem into an interval $[0,R]$, and introduce the linear translation (\ref{linear_trans}) for ${\mathcal V}$ into (\ref{1d_model_cv_1})-(\ref{1d_model_cv_4}) such that both ${\mathcal U}$ and $\overline{\mathcal V}$ satisfy homogeneous Dirichlet boundary conditions. Similar as \cite{ZhangDong}, for computing the dynamics, we first apply the time-splitting technique to decouple the nonlinearity and then use sine pseudospectral method to discretize the spatial derivatives.
Let ${\mathcal U}_j^n\approx {\mathcal U}(r_j, t_n)$ and $\overline{\mathcal V}_j^n\approx \overline{\mathcal V}(r_j,t_n)$.  Choose ${\mathcal U}_j^0={\mathcal U}_0(r_j)$, a second--order time-splitting sine pseudospectral (TSSP) discretization  reads:
\begin{align}
\label{tssp1}& {\mathcal U}_j^{(1)} = \sum_{k=1}^{J-1}\exp\left\{-i\Delta t \mu_k^2/4\right\}\widetilde{\left({\mathcal U}^n\right)}_k \sin \left(\frac{jk\pi}{J}\right),\\
\label{tssp2}& {\mathcal U}_j^{(2)} = \exp\left\{-i\Delta t\left(\Vext(r_j)+\frac{\Cp}{4\pi r_j}\overline{\mathcal V}_j^{(1)}
+\frac{\Cp}{4\pi R} - \alpha \left(2\sqrt{\pi}r_j\right)^{-\frac{2}{3}}\left|{\mathcal U}_j^{(1)}\right|^{\frac{2}{3}}\right)\right\} {\mathcal U}_j^{(1)},\\
\label{tssp3}& {\mathcal U}_j^{n+1} = \sum_{k=1}^{J-1}\exp\left\{-i\Delta t \mu_k^2/4\right\}\widetilde{\left({\mathcal U}^{(2)}\right)}_k \sin \left(\frac{jk\pi}{J}\right),
\intertext{for $n\geq 0$, and $j=1,2,\ldots, J-1$. Here, $\overline{\mathcal V}_j^{(1)}$ is obtained from solving the Poisson equation via sine pseudospectral method (similar as \cite[Section 3.2]{ZhangDong}), i.e.,}
\label{tssp4}& \overline{\mathcal V}_j^{(1)} = \sum_{k=1}^{J-1} \mu_k^{-2}\widetilde{\left(\rho^{(1)}\right)}_k\sin\left(\frac{jk\pi}{J}\right),\quad \mbox{with}\quad
\rho_j^{(1)} = \frac{1}{r_j}\left| {\mathcal U}_j^{(1)}\right|^2,\quad j=1,2.\ldots, J-1.
\end{align}

Again, after we get the solution ${\mathcal U}_j^n$ from (\ref{tssp1})-(\ref{tssp4}), the solution $\psi_j^n\approx \psi(r_j,t_n)$ of (\ref{1d_model_1})-(\ref{1d_model_3}) is achieved via
\begin{equation}
\psi_j^n = \frac{1}{2\sqrt{\pi}}
\left\{
\begin{array}{ll}
{\mathcal U}_j^n/r_j,& j=1,2,\ldots,J,\\
\sum_{k=1}^{J-1} \mu_k \widetilde{\left({\mathcal U}^n\right)}_k,& j=0.
\end{array}
\right.
\end{equation}

The above method is explicit, spectrally accurate in space and second-order accurate in time and it works only when both $\Vext$ and $\psi_0$ are spherically symmetric.  Again, compared with the method proposed in \cite{ZhangDong} for general 3D problem, the memory cost is reduced from $O(J^3)$ to $O(J)$ and computational cost per time step is reduced from $O(J^3\ln(J^3))$ to $O(J\ln(J))$.  In addition, similar as \cite{ZhangDong}, we have,
\begin{lemma}
The TSSP method (\ref{tssp1})-(\ref{tssp4}) is normalization conservation, i.e.,
\begin{equation*}
\left\|{\mathcal U}^n\right\|_h^2: = h_r\sum_{j=1}^{J-1}\left|{\mathcal U}_j^n\right|^2\equiv
h_r\sum_{j=1}^{J-1}\left|{\mathcal U}_j^0\right|^2 =  \left\|{\mathcal U}^0\right\|_h^2,\quad n\geq 0,
\end{equation*}
so it is unconditionally stable in $L^2$-norm.
\end{lemma}

\section{Numerical results}
\setcounter{equation}{0}

\begin{figure}[t!]
\centerline{
\psfig{figure=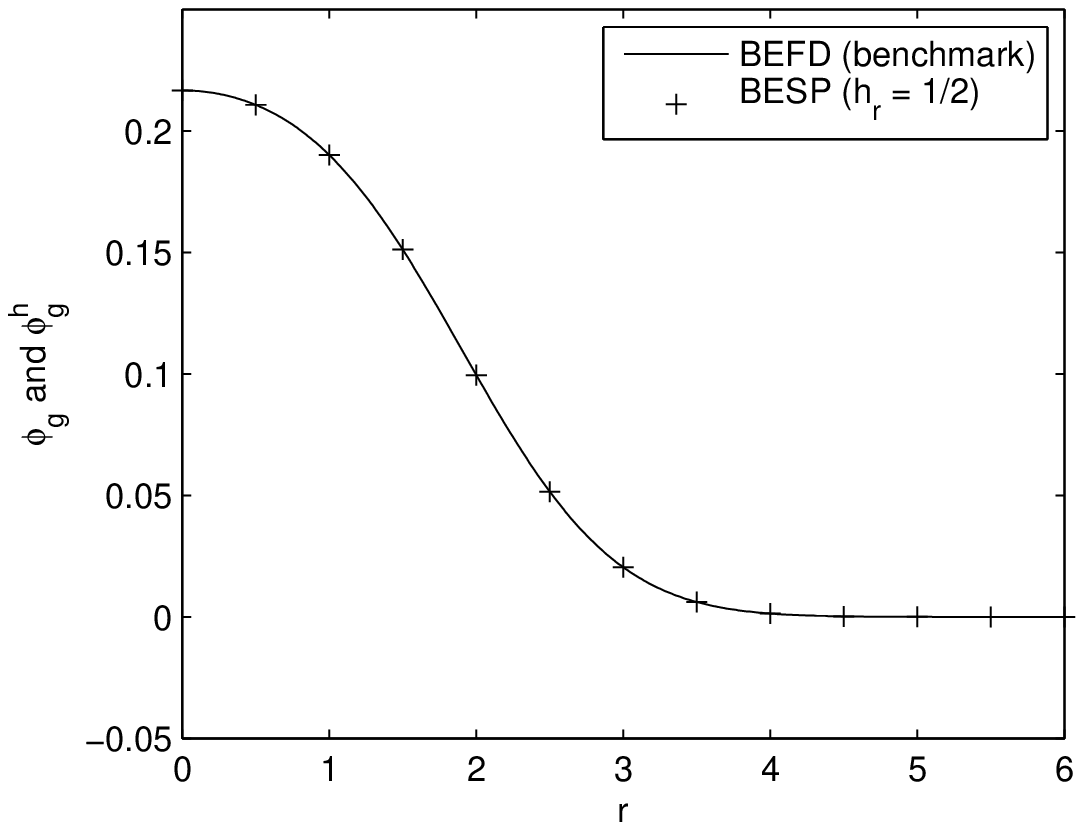,width=8.85cm}
\psfig{figure=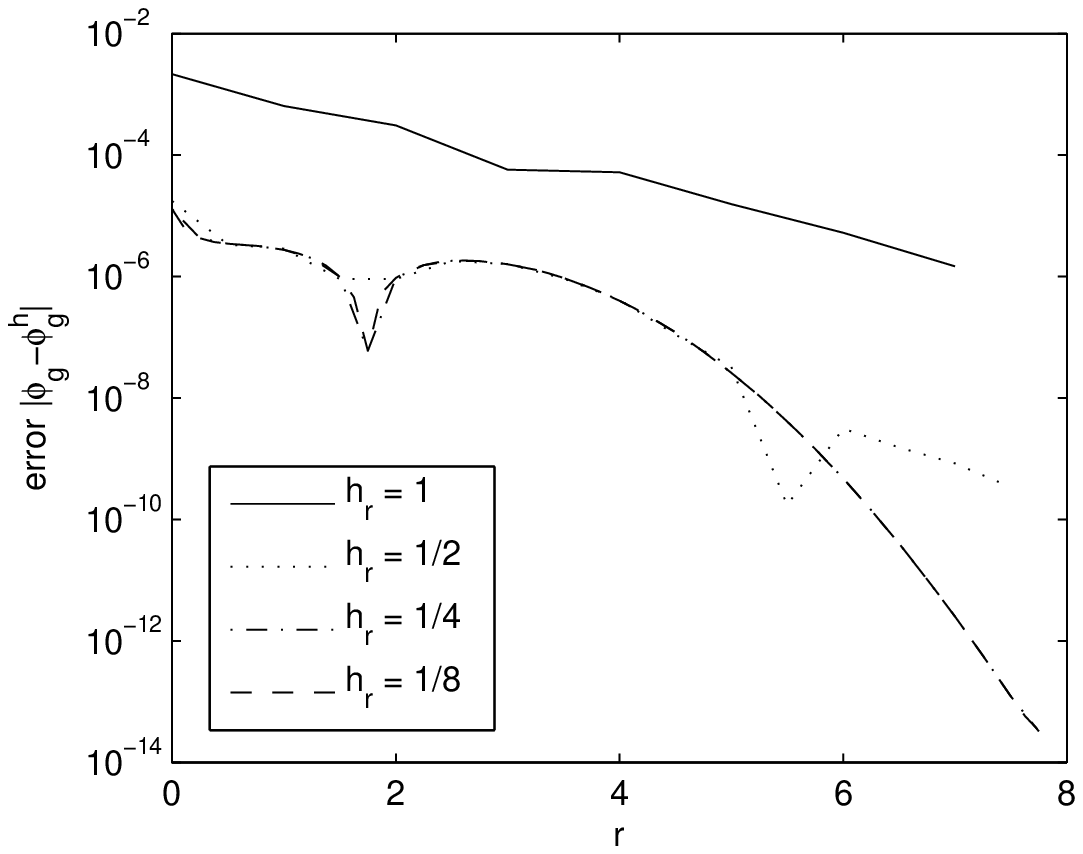,width=8.85cm}}
\caption{Accuracy analysis for BESP method: (1) $\phi_g$ obtained from BEFD method with $h_r=1/64$ as benchmark and $\phi_g^h$ obtained from BESP method with $h_r=1/2$ (left figure); (2) error $\left|\phi_g-\phi_g^h\right|$ with different $h_r$ (right figure).}\label{fig_gs}
\end{figure}

\begin{figure}[t!]
\centerline{
\psfig{figure=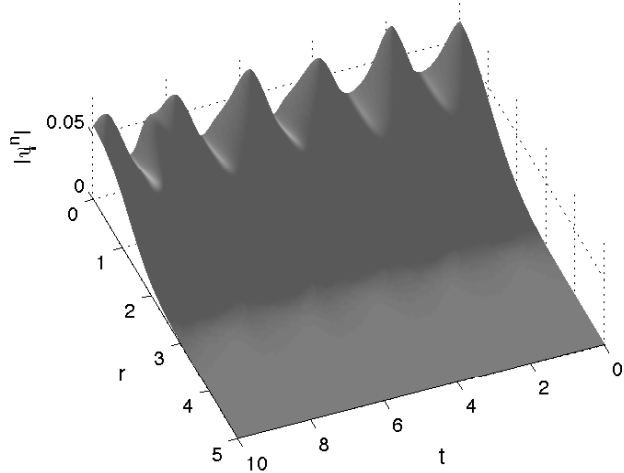,width=8.85cm}}
\caption{Dynamics computed by TSSP method: evolution of $|\psi^n|$ up to time $t_n=10$.}\label{fig_dy}
\end{figure}

Numerical results are reported in this section to demonstrate the accuracy and efficiency of the proposed methods, and we choose $\Vext=\frac{1}{2}r^2$, $\Cp=100$ and $\alpha=1$ in (\ref{1d_model_1}) as the example.  For computing the ground state, the ``exact'' solution $\phi_g$ (benchmark) is achieved by applying a backward Euler finite-difference (BEFD) discretization to a GFDN of the quasi-1D model (\ref{1d_model_1})-(\ref{1d_model_3}) with Dirichlet boundary conditions of $\phi$ and Robin boundary conditions of $\Vp$ \cite{ZhangDong}.  $\phi_g$ is computed in a ball $0\leq r\leq 8$ with a very fine mesh size $h_r=1/64$.  Let $\phi_g^h$ be the approximations obtained from BESP method (\ref{besp1})-(\ref{besp3}), Fig. \ref{fig_gs} plots $\phi_g$ and $\phi_g^h$ with $h_r=1/2$, and the error $\left|\phi_g-\phi_g^h\right|$ with different $h_r$.  The results show that the BESP method (\ref{besp1})-(\ref{besp3}) gives the approximation of ground state with spectral order of accuracy in space; and therefore, it is more efficient in implementation than the standard finite-difference discretization for spherically symmetric case and the spectral-type method proposed in \cite{ZhangDong} for general 3D case.  Similar accuracy and efficiency conclusions can be drawn for TSSP method (\ref{tssp1})-(\ref{tssp3}).  Fig. \ref{fig_dy} plots the evolution of $|\psi^n|$ for $0\leq t_n \leq 10$ when $\psi_0=(2\pi)^{3/4}\exp\left(-{r^2}/{4}\right)$.  Here, the computation is carried out in a ball $0\leq r\leq 16$, with $h_r=1/16$ and $\Delta t=0.01$.

\section{Concluding remarks}

In this study we considered the numerics of the spherically symmetric SPS system and simplified the spectral-type methods proposed in our recent paper \cite{ZhangDong} to compute the ground state and dynamics for general external potential and initial condition.  The simplification is achieved by introducing a proper change of variables into the reduced quasi-1D model.   The simplified methods still admit spectral  order of accuracy in space, with significantly less demand on memory and computational load, and is more efficient in implementation than the standard finite-difference approaches for the spherically symmetric case.  Note that the simplified methods only work for the system with spherical symmetry, and for the general case we still suggest to apply the methods proposed in \cite{ZhangDong}.  Also, the results in this study are applicable to the Schr\"{o}dinger--Poisson and Schr\"{o}dinger--Newton systems ($\alpha=0$ in  (\ref{spsorig})-(\ref{spsorig3})) as well.

\section*{Acknowledgements}
This work was supported by Academic Research Fund of
Ministry of Education of Singapore grant R-146-000-120-112.
Also, the author would like to acknowledge the simulating and helpful discussions with Prof. Weizhu Bao.
Part of this work was done when the author was visiting the
Isaac Newton Institute for Mathematical Sciences in Cambridge.  The visit was supported by the Isaac Newton Institute.

\bigskip


\begin{thebibliography}{10}



\bibitem{BD2011}
W. Bao and X. Dong, Numerical methods for computing ground state and
dynamics of nonlinear relativistic Hartree equation for boson stars,
J. Comput. Phys. 230 (2011), 5449--5469.

\bibitem{BMS}
W. Bao, N.J. Mauser and H.P. Stimming, Effective one particle
quantum dynamics of electrons: A numerical study of the
Schr\"{o}dinger--Poisson--X$\alpha$ model, Comm. Math. Sci. 1 (2003)
809--831.


\bibitem{ADM}
N. Ben Abballan, P. Degond, and P.A. Markowich, On a one-dimensional
Schr\"{o}dinger--Poisson scattering model, ZAMP 48 (1997) 35--55.


\bibitem{BGM}
O. Bokanowski, B. Gr\'{e}bert, and N.J. Mauser, Local density
approximation for the energy of a periodic Coulomb model, Math.
Meth., and Mod. in the Appl. Sci. 13 (8) (2003) 1185--1217.

\bibitem{BLS}
O. Bokanowski, J.L. L\'{o}pez, and J. Soler, On a exchange
interaction model for quantum transport: The
Schr\"{o}dinger--Poisson--Slater system, Math. Model Methods Appl.
Sci. 12 (10) (2003) 1397--1412.

\bibitem{CLLM}
C. Cheng, Q. Liu, J. Lee, and H.Z. Massoud, Spectral element method
for the Schr\"{o}dinger--Poisson system, J. Comput. Electron. 3
(2004) 417--421.


\bibitem{Choquard1}
P. Choquard, J. Stubbe and M. Vuffray, Stationary solutions of the
Schr\"{o}dinger--Newton Model-An ODE approach, Diff. Int. Eqns. 21
(2008) 665--679.

\bibitem{EZ}
M. Ehrhardt and A. Zisowsky, Fast calculation of energy and mass
preserving solutions of Schr\"{o}dinger--Poisson systems on unbounded
domains, J. Comput. Appl. Math. 187 (2006) 1--28.

\bibitem{TodNum2}
R. Harrison, I.M. Moroz and K.P. Tod, A numerical study of
Schr\"{o}dinger--Newton equations,  Nonlinearity, 16 (2003) 101--122.

\bibitem{Kellogg}
O.B. Kellogg, Foundations of potential theory, Dover, New York, 1953.

\bibitem{Lieb}
E.H. Lieb, Existence and uniqueness of the minimizing of Choquards'
nonlinear equation, Studies in Appl. Math. 57 (1976/77) 93--105.


\bibitem{Masaki}
S. Masaki, Energy solution to Schr\"{o}dinger--Poisson system in the
two-dimensional whole space, manuscript.

\bibitem{Mauser}
N.J. Mauser, The Schr\"{o}dinger--Poisson--$X^\alpha$ equation, App.
Math. Letters 14 (2001) 759--763.

\bibitem{Soler}
\'{O}. S\'{a}nchez and J. Soler, Long-Time Dynamics of the
Schr\"{o}dinger--Poisson--Slater Systems, J. Statist. Phys. 114
(2004) 179--204.


\bibitem{Stimming}
H.P. Stimming, The IVP for the Schr\"{o}dinger--Poisson--X$\alpha$
equation in one dimension, Math. Models Methods Appl. Sci. 15
(2005) 1169--1180.

\bibitem{ZhangDong}
Y. Zhang, X. Dong, On the computation of  ground state
and dynamics of Schr\"{o}dinger--Poisson--Slater system, J. Comput. Phys. 230 (2011), 2660--2676.
\end{thebibliography}
\end{document}